\documentclass[superscriptaddress, twocolumn, pre, longbibliography]{revtex4-1}
\usepackage{amsmath, amssymb, color, graphicx, tikz}
\usepackage[utf8]{inputenc}
\definecolor{linkcolor}{rgb}{0,0,0.6} 
\usepackage[pdftex,colorlinks=true,
	pdfstartview = FitV,
	linkcolor    = linkcolor,
	citecolor    = linkcolor,
	urlcolor     = linkcolor,	
	hyperindex   = true,
	hyperfigures = false]{hyperref}

\begin{document}

\title{Quantifying dissipation in flocking dynamics: When tracking internal states matters}

\author{Karel Proesmans}
\affiliation{Niels Bohr International Academy, Niels Bohr Institute, University of Copenhagen, Blegdamsvej 17, 2100 Copenhagen, Denmark}

\author{Gianmaria Falasco}
\affiliation{Department of Physics and Astronomy, University of Padova, Via Marzolo 8, I-35131 Padova, Italy}
\affiliation{INFN, Sezione di Padova, Via Marzolo 8, I-35131 Padova, Italy}

\author{Atul Tanaji Mohite}

\affiliation{Department of Theoretical Physics and Center for Biophysics, Saarland University, Saarbrücken, Germany}
\affiliation{Department of Physics and Materials Science, University of Luxembourg, L-1511 Luxembourg}

\author{Massimiliano Esposito}
\author{\'Etienne Fodor}
\affiliation{Department of Physics and Materials Science, University of Luxembourg, L-1511 Luxembourg}

\begin{abstract}
Aligning self-propelled particles undergo a nonequilibrium flocking transition from apolar to polar phases as their interactions become stronger. We propose a thermodynamically consistent lattice model, in which the internal state of the particles biases their diffusion, to capture such a transition. Changes of internal states and jumps between lattice sites obey local detailed balance with respect to the same interaction energy. We unveil a crossover between two regimes: for weak interactions, the dissipation is maximal, and partial inference (namely, based on discarding the dynamics of internal states) leads to a severe underestimation; for strong interactions, the dissipation is reduced, and partial inference captures most of the dissipation. Finally, we reveal that the macroscopic dissipation, evaluated at the hydrodynamic level, coincides with the microscopic dissipation upon coarse-graining. We argue that this correspondence stems from a generic mapping of active lattice models with local detailed balance into a specific class of non-ideal reaction-diffusion systems.
\end{abstract}

\maketitle


\section{Introduction}

Active matter is a class of nonequilibrium systems where each component individually consumes energy to drive its directed motion~\cite{Marchetti2013, Bechinger2016, Fodor2018}. It encompasses systems made of either synthetic~\cite{Palacci2013, Bartolo2013} or living~\cite{Elgeti2015} self-propelled units. Interactions between active units can lead to collective dynamics without any equivalent in thermal equilibrium, such as the flocking transition to a collective directed motion~\cite{Toner1995, Vicsek1995, Chate2020}, and the motility-induced phase separation (MIPS) of purely repulsive particles~\cite{Cates2015}.

Recently, several works have strived to quantify the departure from equilibrium in models of active matter~\cite{fodor2021irreversibility, Wijland2022}. The motivation is mainly twofold. First, one would like to identify some asymptotic regimes where the dynamics reduces to equilibrium, from which perturbative results can be obtained for the stationary distributions~\cite{Fodor2016, Martin2021} and the irreversibility measures~\cite{Nardini2017, Borthne2020, suchanek2023irreversible, alston2023irreversibility}. Second, one may want to estimate the energy dissipated by the system, as it provides a natural definition of the energy cost required to sustain a given collective phase. Evaluating such a cost is the starting point of any energetic optimization, should it concern either the design of engines~\cite{FodorCates2021, Speck2022} or more generally the control of external perturbations~\cite{davis2023active, alvarado2025}. An important challenge is to formulate and study minimal models which can guide experimental efforts to examine the relations between dissipation and nonequilibrium dynamical patterns~\cite{Foster2021, Foster2023}.

Unambiguously estimating dissipation in active matter is non-trivial. Indeed, active systems feature many degrees of freedom subject to nonequilibrium driving, only a small fraction of which is typically accessible in experiments~\cite{Foster2023}. Therefore, most active models adopt an effective representation of the dynamics, focusing only on a subset of the phase space (e.g., position and orientation of active colloids) and deliberately discarding some underlying processes (e.g., chemical reactions at the colloidal surface). While this standpoint captures the emerging collective phases in many cases~\cite{Marchetti2013, Bechinger2016, Fodor2018}, it overlooks the dissipation of the discarded processes, which can be a large contribution to the total energy cost.

\begin{figure}[b]
	\centering
	\includegraphics[width=.8\linewidth]{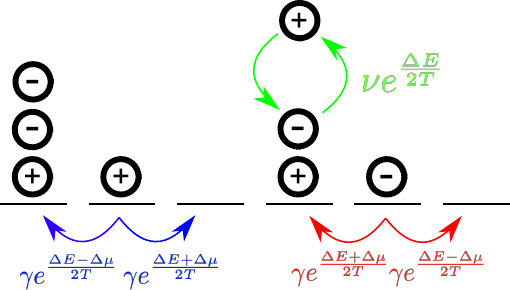}
	\caption{Self-propelled particles $(+,-)$ follow a biased diffusion towards right (blue arrows) and left (red arrows), respectively. Each particle can switch its internal state (green arrows). Local detailed balance enforces that the transition rates are constrained by the energy difference $\Delta E$ between configurations. The temperature $T$ stems from the surrounding thermostat. The chemical potential difference $\Delta\mu$ is fixed by chemostats, and maintains the dynamics arbitrarily far from equilibrium.
	}
	\label{fig:setup}
\end{figure}

Some models have proposed a refined description of active matter, either at particle~\cite{Pietzonka2017, Speck2018, Gaspard2019, Speck2022, Ryabov2022, fritz2023thermodynamically, bebon2024, Golestanian2024, kim2025} or hydrodynamic~\cite{Markovich2021} levels, by describing its coupling to external reservoirs [Fig.~\ref{fig:setup}]. Assuming that the reservoirs are at equilibrium, such a picture is {\it thermodynamically consistent} as it explicitly resolves all nonequilibrium degrees of freedom~\cite{Lebowitz1999, Seifert2012, VandenBroeck2015}. In that respect, active systems amount to passive systems subject to an external drive. For chemically powered activity, the drive is typically the difference between the chemical potentials of reactants and products which fuel the individual self-propulsion. Even when the chemical drive is uniform in space, the resulting driving force need not be, in contrast with other driven systems (e.g., a copper wire subject to an electric field).

While some thermodynamically consistent pictures of MIPS have already been proposed~\cite{Speck2019, bebon2024}, existing models of flocking~\cite{Toner1995, Vicsek1995, Chate2020} do not obey thermodynamically consistency. For instance, the active Ising model (AIM) captures flocking with minimal dynamical rules which can be simulated efficiently~\cite{solon2013revisiting, solon2015flocking}: particles of two types are driven in opposite directions, while aligning with local neighbors [Fig.~\ref{fig:setup}]. This model and its extensions~\cite{chatterjee2020flocking, sakaguchi2019flip, seara2023non, scandolo2023active} have served as a platform to study various aspects of the flocking transition~\cite{solon2022susceptibility, martin2021fluctuation, shen2021interplay} and to benchmark numerical algorithms~\cite{thuroff2014numerical, casert2019interpretable}. Despite its success, the AIM is unsuited for proper thermodynamic analysis, since it does not obey {\em local detailed balance}~\cite{Lebowitz1999, Seifert2012, VandenBroeck2015, Maes}. In fact, although the transitions between particle types are constrained by energetic interactions in AIM, the transitions between sites do not follow any energetic constraint.

In this paper, we introduce a thermodynamically consistent version of the AIM. In Sec.~\ref{sec:model}, we propose a minimal modification of the original AIM~\cite{solon2013revisiting} to obey local detailed balance [Fig.~\ref{fig:setup}] while capturing the three main phases: a homogeneous apolar phase, a coexistence between a polar band and an apolar background, and a homogeneous polar phase. We then evaluate the dissipation in each phase and compare it with two inferred estimations: (i)~the first one accounts for transitions between states and between sites, and (ii)~the second one only accounts for the spatial dynamics, without resolving the particle internal states. The underestimation of dissipation in case (ii) illustrates the limitations of any evaluation that discards some relevant degrees of freedom. In Sec.~\ref{sec:hydro}, we derive a hydrodynamic description of the dynamics in the macroscopic limit, and we demonstrate that the corresponding macroscopic dissipation coincides with its microscopic counterpart. The correspondence between microscopic and macroscopic dissipation relies on a coarse-graining procedure which avoids discarding any relevant contribution to dissipation.

Overall, our results lead to distinguishing two regimes: (i)~for weak interactions (namely, in the apolar phase), the dissipation is maximal, and partial inference (based on discarding the dynamics of internal states) underestimates dissipation, whereas (ii)~for strong interactions (namely, in the polar phase), the dissipation is reduced, and accurately captured by partial inference. Based on thermodynamic consistency, our approach demonstrates how the dissipation of nonequilibrium dynamics can be systematically quantified across scales in various types of active systems, even beyond the case of flocking.


\section{Lattice model of flocking}
\label{sec:model}

We introduce the thermodynamically consistent dynamics of our lattice model [Fig.~\ref{fig:setup}], and demonstrate that the corresponding phase diagram reproduces the three main phases of standard flocking models~\cite{Toner1995, Vicsek1995, Chate2020}. We quantify the entropy production rate in all phases, and compare it with estimations obtained from inference methods.

\subsection{Thermodynamically consistent dynamics}

We consider $N$ particles with internal states $(+,-)$, evolving on a two-dimensional square lattice of size $n\times n$ with periodic boundary conditions. Each lattice site can accommodate an arbitrary number of particles. Particles undergo transitions either between internal states or between lattice sites following a Markov jump process [Fig.~\ref{fig:setup}].

Denoting by $p_{i,j}$ and $m_{i,j}$ the populations of $+$ and $-$ particles at the site $(i,j)$, the state of the system is given by $\alpha=(p_{1,1},m_{1,1},\dots,p_{n,n},m_{n,n})$. The dynamics of the state probability ${\cal P}(\alpha,t)$ reads
\begin{equation}\label{eq:ME}
	\frac{\partial}{\partial t}{\cal P}(\alpha,t) = \sum_{\alpha'} \Big[ {\cal W}_{\alpha,\alpha'}{\cal P}(\alpha',t) - {\cal W}_{\alpha',\alpha}{\cal P}(\alpha,t) \Big] ,
\end{equation}
where ${\cal W}_{\alpha,\alpha'}$ denotes the rate of transition from $\alpha'$ to $\alpha$. Inspired by the original AIM~\cite{solon2013revisiting}, we implement aligning interactions as
\begin{equation}\label{eq:Edef}
	E = - \frac 1 2 {\sum_{i,j}} \bigg[ \varepsilon \frac{P_{i,j}^2}{\rho_{i,j}} + \varepsilon_0 \frac{P_{i,j}}{\sqrt{\rho_{i,j}}}  \bigg( \frac{P_{i+1,j}}{\sqrt{\rho_{i+1,j}}} + \frac{P_{i,j+1}}{\sqrt{\rho_{i,j+1}}} \bigg) \bigg] ,
\end{equation}
where the local density $\rho_{i,j}$ and polarization $P_{i,j}$ read
\begin{equation}\label{eq:density}
	\rho_{i,j} = p_{i,j} + m_{i,j} ,
	\quad
	P_{i,j} = p_{i,j} - m_{i,j} ,
\end{equation}
and $(\varepsilon,\varepsilon_0)>0$ denote the interaction strengths. The functional dependence of the rates ${\cal W}_{\alpha,\alpha'}$ on the energy $E$ differs depending on whether the transition occurs between internal states or between lattice sites [Fig.~\ref{fig:setup}].

On a given lattice site, transitions between internal states follow the rate
\begin{equation}\label{eq:W2}
	\mathcal{W}^{\rm (int)}_{\alpha,\alpha'} = \nu \,e^{ - \frac{ E(\alpha) - E(\alpha')}{2T}} ,
\end{equation}
where $T$ is the temperature of the heat bath (we set the Boltzmann constant to $k_B=1$), and $\nu$ is a constant rate. The energy in Eq.~\eqref{eq:Edef} ensures that the transition rate in Eq.~\eqref{eq:W2} favors transitions towards the state which has locally the largest population. At a fixed internal state, transitions of particles jumping between sites are prescribed by
\begin{equation}\label{eq:W1}
	\mathcal{W}^{\rm (jmp)}_{\alpha,\alpha'} = \gamma \,e^{ - \frac{ E(\alpha) - E(\alpha') + s \Delta\mu}{2T} } ,
\end{equation}
where $\gamma$ is a constant rate. The transitions along $y$ are symmetric ($s=0$), while those along $x$ are biased ($s\neq 0$). In the latter case, the value of $s=\pm 1$ depends on the internal state and on the direction of the jump [Fig.~\ref{fig:setup}]. Thus, the position follows a persistent random walk, where particles tend to preferentially move in the direction set by their internal state.

Our transition rates $\mathcal{W}^{\rm (int)}$ and $\mathcal{W}^{\rm (jmp)}$ follow the thermodynamic constraint of {\it local detailed balance} with respect to the same interaction energy $E$~\cite{Lebowitz1999, Seifert2012, VandenBroeck2015, Maes}. Here, the chemical potential difference $\Delta\mu>0$ represents the energy provided by an underlying chemical reaction which fuels the self-propulsion of particles. We assume that the reaction is tightly coupled to the particle displacement: every product consumption is associated with a biased jump between lattice sites~\cite{Pietzonka2017, Speck2018, fritz2023thermodynamically}.

In the absence of activity ($\Delta\mu=0$), the dynamics reduces to equilibrium: it relaxes to the Boltzmann steady state (${\cal P}_s\sim e^{-E/T}$) analogous to that of an Ising model. The crucial difference with the original AIM~\cite{solon2013revisiting}, and the main common point with another thermodynamically consistent active model~\cite{Agranov2024}, is that the jump rates between sites [Eq.~\eqref{eq:W1}] depend on $E$. While the interaction energy $E$ is defined in Ref.~\cite{Agranov2024} in terms of some mesoscopic variables that are averaged over neighboring lattice sites, we here consider the local variables $(\rho_{i,j},P_{i,j})$ [Eq.~\eqref{eq:density}] in the definition of $E$ [Eq.~\eqref{eq:Edef}]. Yet, in contrast with~\cite{solon2013revisiting}, interactions are not purely local, they now also account for neighboring site populations. As discussed in the following, this choice leads to the emergence of polar bands.



\subsection{Phase diagram: From apolar to polar}

\begin{figure}
  \centering
  \includegraphics[width=\linewidth, trim=0 0 0cm 2cm, clip=true]{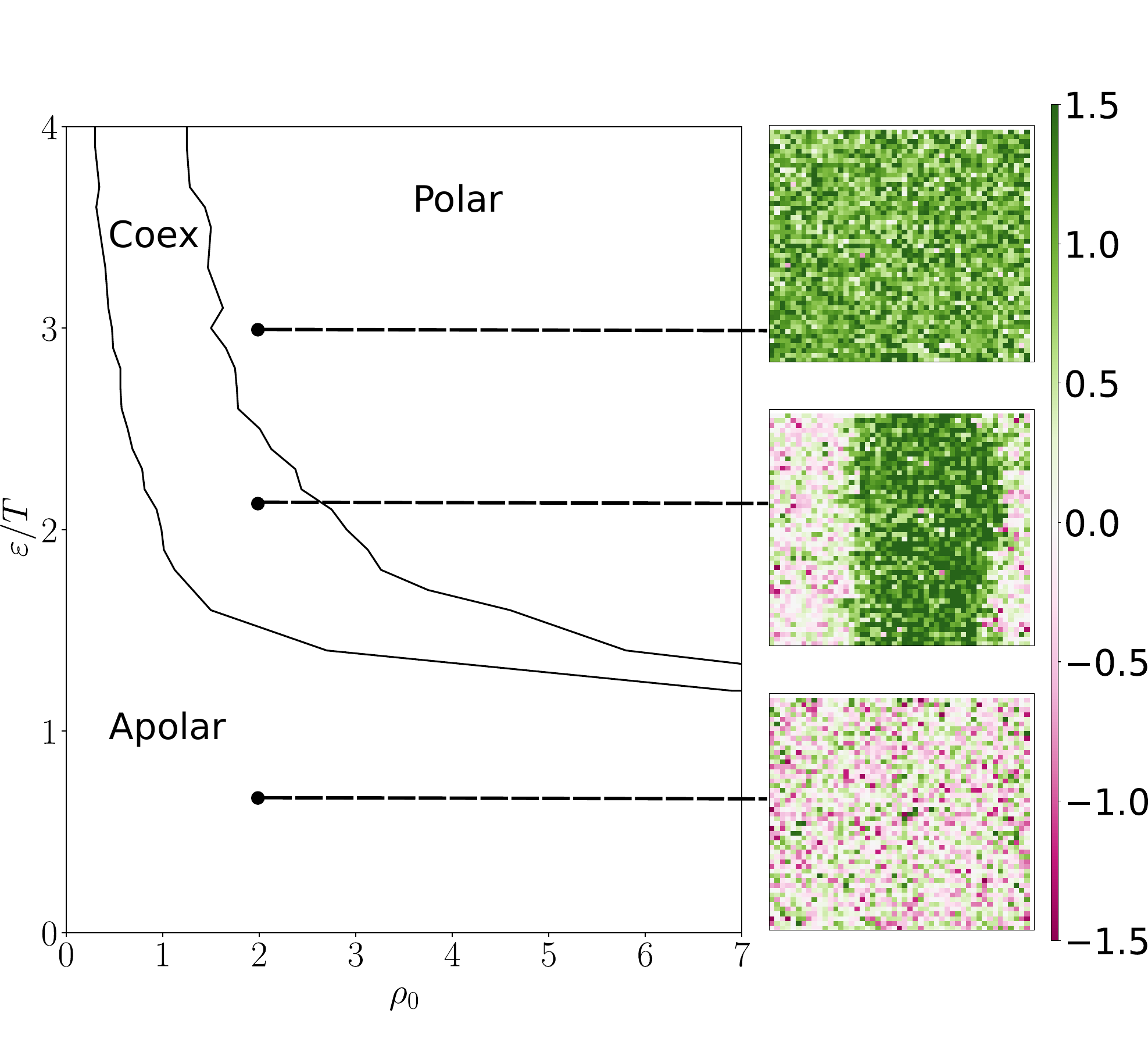}
  \caption{Phase diagram of lattice model as a function of alignment strength $\varepsilon$ and density $\rho_0$. The system is in an apolar phase at small $(\varepsilon, \rho_0)$, and a polar phase at large $(\varepsilon, \rho_0)$. For intermediate values of $(\varepsilon, \rho_0)$, there is coexistence between a dense polar band and a dilute apolar background.
  Snapshots represent the local polarization $P_{i,j}$ [Eq.~\eqref{eq:density}].
  Parameters: $\varepsilon_0=\varepsilon$, $\nu=\gamma=T=1$, $\Delta\mu=0.3$, $n=100$.
  }
  \label{fig:phase-diagram}
\end{figure}

To establish the phase diagram of our dynamics [Eqs.~(\ref{eq:ME}-\ref{eq:W1})], we perform numerical simulations using a Gillespie algorithm~\cite{Gillespie}.  We vary the total density of particles $\rho_0 = N/n^2$ and the interaction strength $\varepsilon$. We identify three different phases analogous to standard flocking models~\cite{Chate2020}: (i)~a disordered apolar phase at low $(\rho_0,\varepsilon)$, (ii)~an ordered polar phase at high $(\rho_0,\varepsilon)$, and (iii)~an intermediate regime of $(\rho_0,\varepsilon)$ where a dense polar band coexists with a dilute apolar background [Fig.~\ref{fig:phase-diagram}]. In particular, the existence of a single polar band is consistent with some related lattice models of flocking, where particles only have two internal states~\cite{solon2013revisiting, Agranov2024}.

In the absence of nearest-neighbor interactions, namely taking $\varepsilon_0=0$ in Eq.~\eqref{eq:Edef}, we do not observe any stable bands. In fact, dense regions with locally high polarization do not convert each other upon collisions, in contrast with the original AIM~\cite{solon2013revisiting}. Instead, such collisions here lead to long-lived configurations where polarized walls moving in opposite directions block each other. Indeed, since transitions between sites now depend on energy differences [Eq.~\eqref{eq:W1}], it is not energetically favorable for a $+$ particle to jump from positively polarized regions to negatively polarized ones (and vice versa for $-$ particles). This kinetic trapping is no longer present in the presence of nearest neighbor interactions.

In short, our model reproduces the main phenomenology of the flocking transition~\cite{Chate2020}. By enforcing local detailed balance in the choice of transition rates [Eqs.~(\ref{eq:W2}-\ref{eq:W1})], we are now in a position to unambiguously define and study the energetics of such a transition.

\subsection{Entropy production rate}\label{sec:dissipation}

The entropy production rate (EPR) $\dot{\cal S}$ associated with our dynamics [Eq.~\eqref{eq:ME}] reads~\cite{Lebowitz1999, Seifert2012}
\begin{equation}\label{eq:sigex}
	\dot{\cal S} = \sum_{\alpha,\alpha'}\mathcal{W}_{\alpha,\alpha'}{\cal P}(\alpha)\ln\left(\frac{\mathcal{W}_{\alpha,\alpha'}{\cal P}(\alpha)}{{\mathcal{W}_{\alpha',\alpha} {\cal P}(\alpha')}}\right) \geq 0. 
\end{equation}
Substituting into Eq.~\eqref{eq:sigex} the expression of the transition rates ${\cal W}_{\alpha,\alpha'}$ [Eqs.~(\ref{eq:W2}-\ref{eq:W1})], we deduce
\begin{equation}\label{eq:epr_micro}
	T\dot{\cal S} = \langle \dot{W}_{\textrm{act}}\rangle - \frac{d}{dt}\left(\left\langle E\right\rangle-T \left\langle S\right\rangle \right).
\end{equation}
The averages of the system's entropy $\langle S\rangle$ and energy $\langle E\rangle$ are given by
\begin{equation}
	\langle S\rangle = \sum_{\alpha}\mathcal{P}(\alpha)\ln\mathcal{P}(\alpha) ,
	\quad
	\langle E\rangle = \sum_{\alpha}\mathcal{P}(\alpha)E(\alpha) ,
\end{equation}
where the microscopic interaction energy $E$ is defined in Eq.~\eqref{eq:Edef}. We have introduced the rate of active work $\langle\dot W_{\rm act} \rangle$, which is produced by jump transitions between sites ${\cal W}^{\rm (jmp)}_{\alpha,\alpha'}$ [Eq.~\eqref{eq:W1}]:
\begin{equation}
	\langle \dot{W}_{\textrm{act}}\rangle = \Delta\mu \sum_{\alpha,\alpha'}\mathcal{W}^{\rm (jmp)}_{\alpha,\alpha'}\mathcal{P}(\alpha') = \Delta\mu \,\langle J_p - J_m\rangle ,
\end{equation}
where $(J_p,J_m)$ denote the rate at which $(+,-)$ particles jump along $\hat{\bf e}_x$, respectively. In what follows, we focus on steady-state statistics, for which 
\begin{equation}\label{eq:epr}
	T \dot{\cal S} = \langle \dot{W}_{\textrm{act}}\rangle_{\rm ss} = \Delta\mu \,\langle J_p - J_m\rangle_{\rm ss} .
\end{equation}
where $\langle\cdot\rangle_{\rm ss}$ denotes the steady-state average. The stationary EPR, also referred to as the {\em dissipation rate}, quantifies the rate at which energy is dissipated to drive the motility of $(+,-)$ particles. It vanishes in the absence of activity ($\Delta\mu=0$), as expected for thermodynamically consistent dynamics. In contrast, computing EPR in the original AIM and its variants~\cite{solon2013revisiting, yu2022energy} does not lead to vanishing dissipation when the particles are not self-propelled.

\begin{figure}
  \centering
  \includegraphics[width=.85\linewidth]{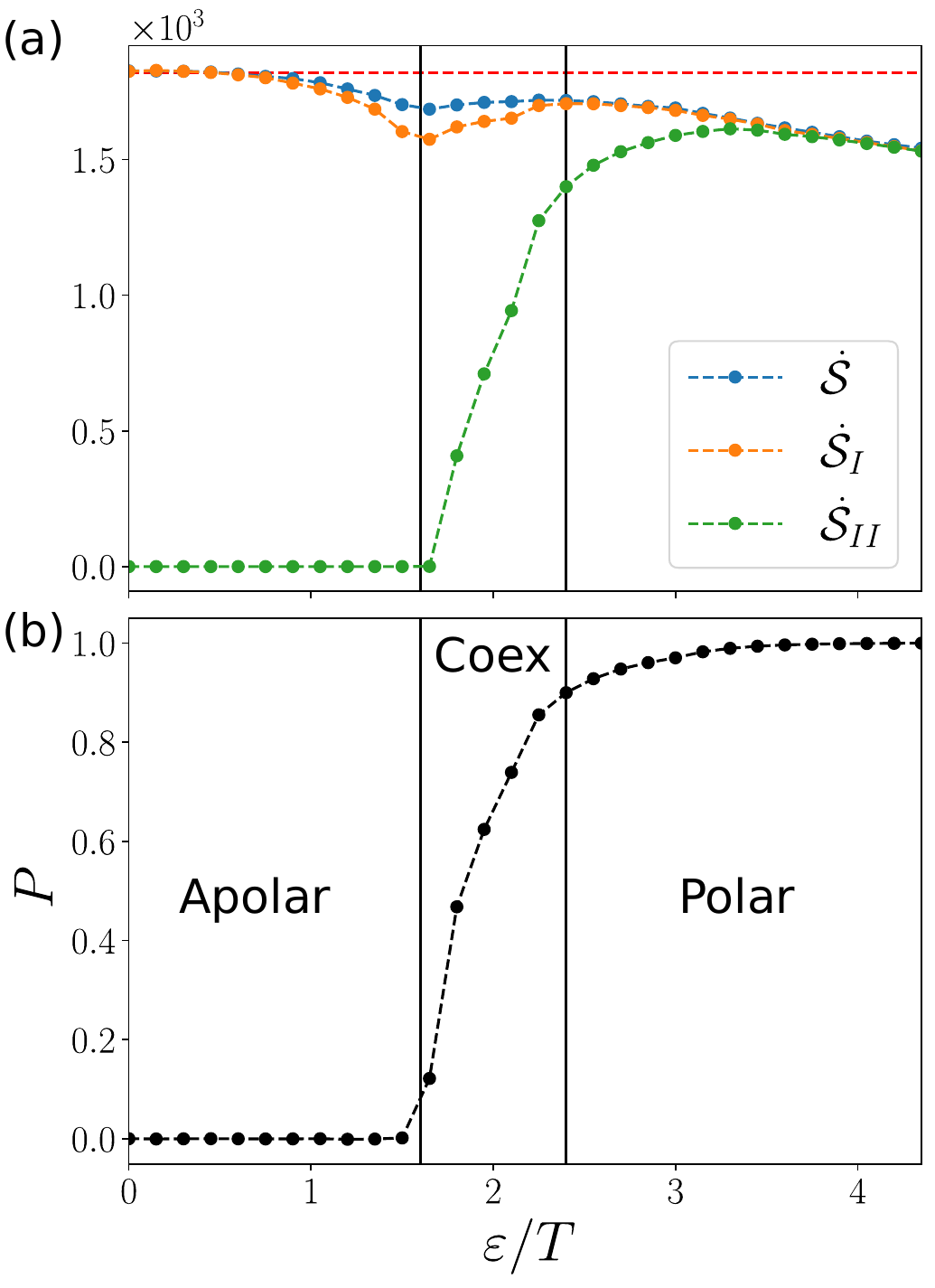}
  \caption{(a)~Entropy production rate (EPR) per particle, and (b)~global polarization $\bar P = \frac{1}{N} \sum_{i,j}P_{i,j}$ as functions of alignment strength $\varepsilon/T$. In panel (a), we compare the direct evaluation of EPR [$\dot{\cal S}$, Eq.~\eqref{eq:sigex}] with estimation from two inference methods, which either track [$\dot{\cal S}_{\rm I}$, Eq.~\eqref{eq:sigsp1}] or discard [$\dot{\cal S}_{\rm II}$, Eq.~\eqref{eq:sigsp2}] internal states. The dashed red line refers to the limit for non-interacting particles [Eq.~\eqref{eq:epr_lim}].
  Parameters: same as in Fig.~\ref{fig:phase-diagram}, with $\rho_0=2$.
  }
  \label{fig:dissipation}
\end{figure}

We numerically evaluate the dissipation rate $\dot{\cal S}$ [Eq.~\eqref{eq:epr}] as a function of the alignment strength $\varepsilon$  [Fig.~\ref{fig:dissipation}]. At small $\varepsilon$, the dissipation rate is highest and coincides with the value for non-interacting particles:
\begin{equation}\label{eq:epr_lim}
	T \dot{\cal S} \underset{\varepsilon\to0}{\longrightarrow} N \gamma \frac{\Delta \mu^2}{T} .
\end{equation}
Moreover, $\dot{\cal S}$ decreases with $\varepsilon$ in the apolar and polar phases. For the intermediate regime of $\varepsilon$ where a polar band forms, instead $\dot{\cal S}$ increases with $\varepsilon$. Our results are in stark contrast with the evaluation of EPR in the original AIM~\cite{yu2022energy}, which yields a cusp at the transition between apolar and coexistence phases. In fact, such a cusp stems from discarding the dependence of ${\cal W}^{\rm (jmp)}_{\alpha,\alpha'}$ on $E$ [Eq.~\eqref{eq:W1}]: it remains present even in the absence of self-propulsion~\cite{yu2022energy}, showing that it is due to the lack of thermodynamic consistency.

\subsection{Inference from tracer dynamics}

Evaluating the dissipation rate $\dot{\cal S}$ [Eq.~\eqref{eq:epr}] requires tracking the internal states of particles. In fact, measuring the currents $(J_p,J_m)$ assumes that one can distinguish $(+,-)$ particles. Yet, in many experiments of self-propelled particles, although one can typically track the displacement of particles, it is often difficult to determine the self-propulsion force (which does not coincide with velocity a priori). An open question is: how accurately can one estimate EPR despite the inability to track internal states? To examine this issue, we consider some strategies for evaluating EPR with only a partial knowledge of the dynamics.




\subsubsection{Tracking internal states}

Inspired by~\cite{lynn2022decomposing, lynn2022emergence}, we explore under which conditions the EPR can be inferred from the dynamics of a tracer embedded in the system. Specifically, our aim is to reduce our many-body dynamics [Eq.~\eqref{eq:ME}] into the dynamics of some representative particles for each state $(+,-)$. To this end, we consider the rate $j_{s\uparrow}$ at which particles in a given state $s\in(+,-)$ jump to the up-site (namely, above its present location); similarly, we introduce the rates $(j_{s\downarrow}, j_{s\leftarrow}, j_{s\rightarrow})$ for jumps in other directions.

\begin{figure}
  \centering
  \includegraphics[width=\linewidth, trim=1cm 0cm 2cm 0, clip=ture]{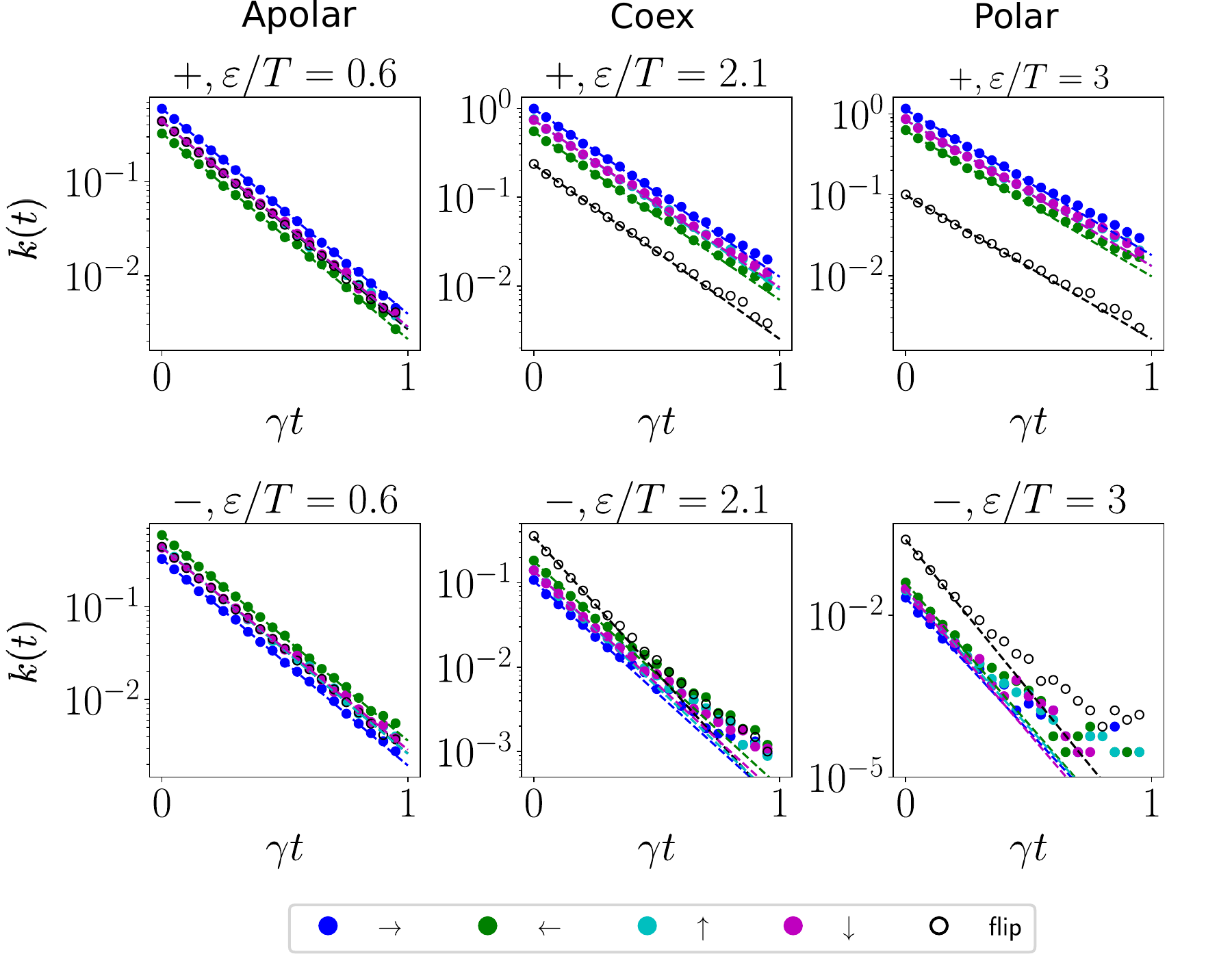}
  \caption{Probability distribution $k(t)$ of the waiting time $t$ between transitions where particles either jump between lattice sites ($\uparrow, \downarrow, \leftarrow, \rightarrow$) or change their internal state (flip). We distinguish (top)~$+$ and (bottom)~$-$ particles in various phases: (left)~apolar, (center)~coexistence, and (right)~ polar. The dashed lines are exponential fits.
  Same parameters as in Fig.~\ref{fig:dissipation}. Note that the polar and coexistence phases preferentially feature $+$ particles due to initially conditions, which explains the asymmetry in the distributions $p(t)$ for $(+,-)$ particles.
  }
  \label{fig:waitingtime}
\end{figure}

The jump rates can be numerically evaluated from the statistics of waiting times (namely, the time between two subsequent transitions). The waiting times essentially follow Poisson distributions, with some deviations in the distribution tails for the polar phase [Fig.~\ref{fig:waitingtime}]. These results suggest that tracers obey a Markovian dynamics to a good approximation: we then extract $(j_{s\uparrow}, j_{s\downarrow}, j_{s\leftarrow}, j_{s\rightarrow})$ as the inverse of the corresponding average waiting time.

Our evaluation of the rates assumes that each $(+,-)$ particle follows a similar dynamics independently of its location. In other words, we deliberately discard any effects stemming from spatial inhomogeneities. The EPR of the tracer dynamics then follows as
\begin{equation}\label{eq:sigsp1}
	\dot{\cal S}_{\rm I} = \sum_{s\in(+,-)} \bigg[ (j_{s\uparrow}-j_{s\downarrow}) \ln\frac{j_{s\uparrow}}{j_{s\downarrow}} + (j_{s\leftarrow}-j_{s\rightarrow}) \ln\frac{j_{s\leftarrow}}{j_{s\rightarrow}} \bigg] .
\end{equation}
The inferred EPR $\dot{\cal S}_{\rm I}$ [Eq.~\eqref{eq:sigsp1}] can be regarded as a coarse-grained version~\cite{Esposito2012} of the original EPR $\dot{\cal S}$ [Eq.~\eqref{eq:sigex}], and it should match $\dot{\cal S}$ whenever interactions are negligible~\cite{lynn2022decomposing}. Our numerical estimation shows that $\dot{\cal S}_{\rm I} \simeq \dot{\cal S}$ not only in the apolar phase, as expected, but also in the polar state [Fig.~\ref{fig:dissipation}]. In fact, slight deviations are observed close to the coexistence phase, namely when spatial inhomogeneities (which are discarded by our inference scheme) affect the dynamics.

\subsubsection{Discarding internal states}

Assuming that one cannot track internal states, the tracer dynamics is then prescribed by effective transition rates which do not distinguish $(+,-)$ particles. The inferred EPR correspondingly reads
\begin{equation}\label{eq:sigsp2}
	\dot{\cal S}_{\rm II} = (j_{\uparrow}-j_{\downarrow}) \ln\frac{j_{\uparrow}}{j_{\downarrow}} + (j_{\leftarrow}-j_{\rightarrow}) \ln\frac{j_{\leftarrow}}{j_{\rightarrow}} ,
\end{equation}
where 
\begin{equation}
	j_\uparrow = \sum_{s\in(+,-)} j_{s\uparrow} ,
\end{equation}
and a similar definition holds for $(j_{\downarrow}, j_{\leftarrow}, j_{\rightarrow})$. The inferred EPR $\dot{\cal S}_{\rm II}$ [Eq.~\eqref{eq:sigsp2}] can be regarded as a coarse-grained version of $\dot{\cal S}_{\rm I}$ [Eq.~\eqref{eq:sigsp1}]. One can show that~\cite{Esposito2012, VandenBroeck2015}
\begin{equation}
	0 \leq \dot{\cal S}_{\rm II} \leq \dot{\cal S}_{\rm I} \leq \dot{\cal S} . 
\end{equation}
Therefore, the partial knowledge of the dynamics generally leads to underestimating the EPR.

The deviation of $\dot{\cal S}_{\rm II}$ from $\dot{\cal S}$ is most striking in the apolar phase, and progressively reduces as $\varepsilon$ increases, until $\dot{\cal S}_{\rm II} \simeq \dot{\cal S}$ deep in the polar phase [Fig.~\ref{fig:dissipation}]. In fact, the evolution of $\dot{\cal S}_{\rm II}$ is qualitatively (but not quantitatively) similar to that of global polarization $\bar P = \frac{1}{N}\sum_{i,j}P_{i,j}$. Interestingly, $\dot{\cal S}_{\rm II}$ vanishes in the apolar phase: this result suggests that the effective tracer dynamics may appear as an effective equilibrium dynamics in this regime, provided that one does not track the internal states.


\subsubsection{What do we learn from inference?}

Overall, our results show that the inferred EPR $\dot{\cal S}_{\rm I}$ [Eq.~\eqref{eq:sigsp1}] is rather close to the original $\dot{\cal S}$ [Eq.~\eqref{eq:sigex}] in all phases: when tracking internal states, our inference method provides a satisfactory estimation. Instead, a significant discrepancy arises when discarding internal states [$\dot{\cal S}_{\rm II}$, Eq.~\eqref{eq:sigsp2}]. By definition, $\dot{\cal S}_{\rm II}$ only detects the dissipation resulting from particle displacement: it discards any contributions from the dynamics of internal states, which are dominant in the apolar phase and negligible in the polar phase.

In short, the flocking transition is associated with a crossover between two regimes: (i)~at weak interactions (namely, low $\varepsilon$ and $\rho_0$), partial inference (that discards the dynamics of internal states) predicts a vanishing EPR, although the EPR is actually very high, whereas (ii)~at strong interactions (namely, high $\varepsilon$ and $\rho_0$), partial inference accurately captures the actual EPR. Interestingly, the dissipation rate is highest (and identical to the value for non-interacting active particles) in the apolar phase, namely in the absence of any collective effects.


\section{Hydrodynamics of flocking}\label{sec:hydro}

We obtain the hydrodynamic description of our lattice model [Fig.~\ref{fig:setup}] in the macroscopic limit of large system sizes and small lattice spacing where we discard field fluctuations~\cite{falasco2023macroscopic}. We compute the corresponding EPR, and demonstrate that it coincides with that of our lattice model.


\subsection{Macroscopic field dynamics}

We introduce the fields $(p,m)$ defined by
\begin{equation}\label{eq:field_def}
	p({\bf r}_{ij},t) = p_{i,j}(t) ,
	\quad
	m({\bf r}_{ij},t) = m_{i,j}(t) ,
\end{equation}
where ${\bf r}_{ij} = a (i \hat{\bf e}_x+ j \hat{\bf e}_y)$, and $a$ is the lattice spacing. The fields of local density $\rho$ and polarization $P$ [Eq.~\eqref{eq:density}] are simply related to $(p,m)$ as
\begin{equation}\label{eq:corr}
	p=(\rho+P)/2,
	\quad
	m=(\rho-P)/2.
\end{equation}
The interaction energy $E$ [Eq.~\eqref{eq:Edef}] can be written in terms of these fields as
\begin{equation}\label{eq:E_hydro}
	E = - \frac{1}{2} \int_V d{\bf r} \Bigg[ \varepsilon \frac{P^2}{\rho} - \bar\varepsilon \bigg(\nabla \frac{P}{\sqrt \rho} \bigg)^2 \Bigg] ,
\end{equation}
where $V$ is the system size, and
\begin{equation}
	\bar\varepsilon = \lim_{a\to0} (\varepsilon_0 a^2).
\end{equation}
Using path-integral methods, we obtain the equations for the most likely value of the fields $(p,m)$, which we denote by the same symbols as the fluctuating fields for convenience [Appendix~\ref{app:action}], as
\begin{equation}\label{eq:hydro}
\begin{aligned}
  \partial_t p &= D \nabla \cdot \left[ \frac{p}{T} \left( \nabla  \frac{\delta E}{\delta p} + {\bf f} \right) + \nabla p \right] - p {\cal W}_+ + m {\cal W}_- ,
  \\
  \partial_t m &= D \nabla \cdot \left[ \frac{m}{T} \left( \nabla  \frac{\delta E}{\delta m} - {\bf f} \right) + \nabla m \right] + p {\cal W}_+ - m {\cal W}_- ,
  \end{aligned}
\end{equation}
where the diffusion coefficient $D$ and the force $\bf f$ read
\begin{equation}\label{eq:f_scale}
	D = \lim_{a\to0} (\gamma a^2) ,
	\quad
	{\bf f} = \hat{\bf e}_x \lim_{a\to 0} (\Delta\mu/a) ,
\end{equation}	
and we have introduced the rates
\begin{equation}\label{eq:rate_scale}
	{\cal W}_{+} = \nu e^{\frac{1}{2T} \left( \frac{\delta E}{\delta p} - \frac{\delta E}{\delta m} \right)} ,
	\quad
	{\cal W}_{-} = \nu e^{\frac{1}{2T} \left( \frac{\delta E}{\delta m} - \frac{\delta E}{\delta p} \right)} .
\end{equation}
Therefore, the hydrodynamic equations for $(p,m)$ are readily written in a closed form.

Such hydrodynamic equations are analogous to those of a non-ideal reaction-diffusion system~\cite{Agranov2024, Avanzini2024}. In fact, our lattice model describes a dynamics with reactions [green arrows in Fig.~\ref{fig:setup}] and biased diffusion [red and blue arrows in Fig.~\ref{fig:setup}] for a mixture of $(+,-)$ interacting particles. The thermodynamic consistency of the lattice dynamics ensures that the corresponding hydrodynamics maps into equilibrium in the absence of activity (${\bf f}\to{\bf 0}$). In this regime, one can show that the equilibrium free energy
\begin{equation}\label{eq:F}
	F = E + T \int_V d{\bf r} ( p \ln p + m \ln m ) 
\end{equation}
is the Lyapunov function which determines the stationary properties~\cite{Aslyamov2023, Agranov2024, Avanzini2024}, as expected for passive dynamics.

Using Eq.~\eqref{eq:corr}, we readily deduce the closed form of hydrodynamic equations for $(\rho,P)$ from Eqs.~(\ref{eq:hydro}-\ref{eq:rate_scale}) as 
\begin{equation}\label{PDEs_in}
  \partial_t \rho = - \nabla \cdot {\bf j}_\rho
  \quad
  \partial_t P = - \nabla \cdot {\bf j}_P + \sigma_P ,
\end{equation}
where the diffusive currents $({\bf j}_\rho, {\bf j}_P)$ read
\begin{equation}\label{eq:currents}
\begin{aligned}
  {\bf j}_\rho &= - \frac{D}{T} \left( \rho \nabla \frac{\delta E}{\delta \rho} +P \nabla  \frac{\delta E}{\delta P} + P {\bf f} \right) + D \nabla \rho ,
	\\
	{\bf j}_P &= - \frac{D}{T} \left( P \nabla  \frac{\delta E}{\delta \rho} + \rho \nabla  \frac{\delta E}{\delta P} + \rho {\bf f} \right) + D \nabla P ,
\end{aligned}
\end{equation}
and the source term $\sigma_P$ is given by 
\begin{equation}
	\sigma_P = 2 \gamma \left[\rho \sinh \left(-\frac 1 T \frac{\delta E}{\delta P} \right) - P \cosh \left(-\frac 1 T \frac{\delta E}{\delta P} \right)\right] .
\end{equation}
Setting $E=0$ in the diffusive currents [Eq.~\eqref{eq:currents}], which amounts to assuming that transitions between microscopic sites ${\cal W}^{\rm (jmp)}_{\alpha,\alpha'}$ [Eq.~\eqref{eq:W1}] are independent of interactions, one recovers the hydrodynamic equations of the original AIM~\cite{solon2013revisiting}. In our case, the explicit dependence of $({\bf j}_\rho, {\bf j}_P)$ and $\sigma_P$ on interactions is the defining feature of non-ideal reaction-diffusion dynamics.

The fixed points $(\rho_0, P_0)$ obey $\sigma_P=0$, since the diffusive currents vanish for homogeneous phases, yielding
\begin{equation} 
 \frac{P_0}{\rho_0} = \tanh\bigg(-\frac 1 T \frac{\delta E}{\delta P} \Big|_{\rho_0, P_0} \bigg) = \tanh\bigg(\frac{\varepsilon}{T} \frac{P_0}{\rho_0} \bigg) ,
\end{equation}
where we have used the expression of the interaction energy $E$ [Eq.~\eqref{eq:E_hydro}]. As in the original AIM~\cite{solon2013revisiting}, we deduce that there exist two solutions: (i)~an apolar phase ($P_0=0$) at small $(\rho_0, \varepsilon)$, and (ii)~a polar phase ($P_0>0$) at large $(\rho_0, \varepsilon)$. Linear stability analysis shows that both phases are always stable whenever $\bar\varepsilon=0$ in $E$ [Eq.~\eqref{eq:E_hydro}], and we expect this result to hold even in the presence of nearest-neighbor interactions.

In short, by coarse-graining our lattice model, we obtain a thermodynamically consistent description at the hydrodynamic level, which maps into a non-ideal reaction diffusion system~\cite{Aslyamov2023, Avanzini2024}. In the macroscopic regime where we neglect fluctuations, the system only adopts a homogeneous configuration, either apolar or polar, in contrast with our lattice model where a traveling band can emerge [Fig.~\ref{fig:phase-diagram}]. In fact, accounting for weak hydrodynamic fluctuations would lead to renormalizing hydrodynamic coefficients~\cite{martin2021fluctuation}, which could be sufficient to destabilize homogeneous phases and yield a traveling band. Instead, for the thermodynamically consistent model in Ref.~\cite{Agranov2024}, we expect that fluctuations do not renormalize the hydrodynamic equations since microscopic interactions depend on some mesoscopic variables (namely, averaged over neighboring sites).


\subsection{Macroscopic entropy production rate}

To evaluate the macroscopic EPR, we cast the macroscopic field dynamics [Eq.~\eqref{eq:hydro}] in a convenient form~\cite{Aslyamov2023, Avanzini2024}
\begin{equation}\label{eq:dyn_hydro}
  \partial_t p = - \nabla \cdot {\bf j}_p + \sigma ,
  \quad
  \partial_t m = - \nabla \cdot {\bf j}_m - \sigma ,
\end{equation}
where the source term $\sigma$ is given by
\begin{equation}\label{eq:reac_rate}
	\sigma = m {\cal W}_- - p {\cal W}_+ ,
\end{equation}
and the diffusive currents $({\bf j}_p,{\bf j}_m)$ read
\begin{equation}\label{eq:diff_current}
	{\bf j}_p = - \mu_p \bigg( \nabla \frac{\delta F}{\delta p} + {\bf f} \bigg) ,
	\quad
	{\bf j}_m = - \mu_m \bigg( \nabla \frac{\delta F}{\delta m} - {\bf f} \bigg) ,
\end{equation}
with the free energy $F$ [Eq.~\eqref{eq:F}] and the mobilities
\begin{equation}
	\mu_p = \frac D T p ,
	\quad
	\mu_m = \frac D T m .
\end{equation}
The EPR associated with this hydrodynamics can be decomposed as
\begin{equation}\label{eq:epr_macro}
	\dot{\cal S}_{\rm macro} = \dot{\cal S}_{\rm diff} + \dot{\cal S}_{\rm reac} ,
\end{equation}
where the contributions from the diffusive ($\dot{\cal S}_{\rm diff}$) and reactive ($\dot{\cal S}_{\rm reac}$) sectors of the dynamics read
\begin{equation}\label{eq:epr_macro_bis}
\begin{aligned}
	T\dot{\cal S}_{\rm diff} &= \int_V \bigg( \frac{{\bf j}_p^2}{\mu_p} + \frac{{\bf j}_m^2}{\mu_m} \bigg) d{\bf r} ,
	\\
	T\dot{\cal S}_{\rm reac} &= \int_V \sigma \bigg( \frac{\delta F}{\delta m} - \frac{\delta F}{\delta p} \bigg) d{\bf r} .
\end{aligned}
\end{equation}
Such a decomposition of EPR is generic for thermodynamically consistent dynamics of non-ideal reaction-diffusion; for instance, see Sec.~IV.B in~\cite{Avanzini2024}. In the presence of noise terms, evaluating the irreversibility of field fluctuations $(p,m)$ leads to the same expression of EPR~\cite{Agranov}, as expected.

We now demonstrate that the macroscopic EPR [$\dot{\cal S}_{\rm macro}$, Eqs.~(\ref{eq:epr_macro}-\ref{eq:epr_macro_bis})] coincides with the macroscopic limit of the microscopic EPR [$\dot{\cal S}$, Eq.~\eqref{eq:epr}]. To this end, we evaluate how the free energy $F$ varies in time: 
\begin{equation}
\begin{aligned}
	\frac{d F}{d t} &= \int_V \bigg[ \frac{\delta F}{\delta p} \partial_t p + \frac{\delta F}{\delta m} \partial_t m \bigg] d{\bf r}
	\\
	&= \int_V \bigg[ \sigma \bigg( \frac{\delta F}{\delta p} - \frac{\delta F}{\delta m} \bigg) + {\bf j}_p \cdot \nabla \frac{\delta F}{\delta p} + {\bf j}_m \cdot \nabla \frac{\delta F}{\delta m} \bigg] d{\bf r} ,
\end{aligned}
\end{equation}
where we have used the dynamics [Eq.~\eqref{eq:dyn_hydro}], and integrated by parts. Expressing the chemical potentials $(\frac{\delta F}{\delta p}, \frac{\delta F}{\delta m})$ in terms of the diffusive currents [Eq.~\eqref{eq:diff_current}], we deduce
\begin{equation}\label{eq:epr_macro_split}
	T \dot{\cal S}_{\rm macro} =  {\bf f} \cdot \int_V ({\bf j}_p - {\bf j}_m) d{\bf r}
    - \frac{d F}{dt},
\end{equation}
where we have used the definition of $\dot{\cal S}_{\rm macro}$ [Eqs.~(\ref{eq:epr_macro}-\ref{eq:epr_macro_bis})]. The connection in Eq.~\eqref{eq:epr_macro_split} between the macroscopic EPR $\dot {\cal S}_{\rm hydro}$, the free energy $F$, and the non-conservative force ${\bf f}$ mirrors the equivalent version for the lattice model [Eq.~\eqref{eq:epr_micro}]. These generic relations hold for an arbitrary interaction energy $E$.

In steady state, the contribution $\frac{d F}{d t}$ vanishes in Eq.~\eqref{eq:epr_macro_split}. In the macroscopic regime, where field fluctuations can be neglected, the relation
\begin{equation}
	\Delta\mu \,\langle J_p - J_m \rangle_{\rm ss} \longrightarrow {\bf f} \cdot \int_V ({\bf j}_p - {\bf j}_m) d{\bf r} ,
\end{equation}
together with Eq.~\eqref{eq:epr}, then leads to the correspondence between macroscopic and microscopic EPR:
\begin{equation}
	\dot{\cal S} \longrightarrow \dot{\cal S}_{\rm macro} .
\end{equation}
Noting that ${\bf j}_p - {\bf j}_m= {\bf j}_P$, and using the explicit expression of the diffusive current ${\bf j}_P = - \frac D T \rho_0 {\bf f}$ [Eq.~\eqref{eq:currents}] for the homogeneous stationary profile $(\rho,P)=(\rho_0,P_0)$, we deduce 
\begin{equation}\label{eq:epr_2}
	T \dot{\cal S}_{\rm macro} = \frac D T \rho_0 V {\bf f}^2 .
\end{equation}
Therefore, within our macroscopic field dynamics, the EPR is constant in all phases (apolar and polar), and coincides with the value for a non-interacting system ($\varepsilon=0$). Using $\lim_{a\to 0} (\gamma\Delta\mu^2) = D {\bf f}^2$ [Eq.~\eqref{eq:f_scale}] and $\rho_0 V = N$, we recover the expression of the microscopic EPR [Eq.~\eqref{eq:epr_lim}] in the regime of weak interactions, namely small $(\varepsilon,\rho_0)$, as expected.

In short, the thermodynamically consistent formulation of the lattice model and its hydrodynamic description entails a correspondence between the EPR evaluated at the microscopic and macroscopic levels. Since our hydrodynamics only features a homogeneous steady state, it does not reproduce the variation of EPR reported in the lattice model [Fig.~\ref{fig:dissipation}].

We can propose a macroscopic equivalent of the microscopic EPR inferred without tracking internal states [$\dot{\cal S}_{\rm II}$, Eq.~\eqref{eq:sigsp2}]. In fact, such an inference amounts to assuming that one has only access to the dynamics of the density $\rho$, namely without tracking the polarization $P$. Since the density has a purely diffusive dynamics [Eq.~\eqref{PDEs_in}], with mobility $\frac D T \rho$ and current ${\bf j}_\rho$ [Eq.~\eqref{eq:currents}], the associated EPR reads
\begin{equation}
	T\dot{\cal S}_{\rm II,macro} = \frac T D \int_V \frac{{\bf j}_\rho^2}{\rho} d{\bf r} .
\end{equation}
In steady state, we get ${\bf j}_\rho = - \frac D T P_0 {\bf f}$ [Eq.~\eqref{eq:currents}], yielding
\begin{equation}
	T \dot{\cal S}_{\rm II,macro} = \frac D T \frac{P_0^2}{\rho_0} V {\bf f}^2 .
\end{equation}
Therefore, estimating the EPR inferred without tracking internal states amounts to replacing $\rho_0$ by $P_0^2/\rho_0$ in the original EPR [Eq.~\eqref{eq:epr_2}]. Interestingly, $\dot{\cal S}_{\rm II,macro}$ varies with interactions through dependence on polarization $P_0$, in contrast to $\dot{\cal S}_{\rm macro}$. In fact, such a dependence is in qualitative agreement with the numerical estimations in our lattice dynamics: it vanishes in the apolar phase, and continuously increases with $\varepsilon$ in the coexistence and polar phases; see the polarization $P$ and inferred EPR $\dot{\cal S}_{\rm II}$ in Fig.~\ref{fig:dissipation}.


\section{Discussion}

We have examined how to unambiguously quantify the dissipation in a lattice model of flocking. Inspired by the original AIM~\cite{solon2013revisiting, solon2015flocking}, our model features two species of particles which (i)~self-propel in opposite directions, and (ii)~change the internal state that determines this direction. Crucial to our approach is {\it local detailed balance}: it ensures that transitions between states and jumps between lattice sites are constrained by the same interaction energy~\cite{Lebowitz1999, Seifert2012, VandenBroeck2015, Maes}. This constraint leads to a correspondence between the EPR evaluated at the microscopic and macroscopic levels.

We demonstrate that the EPR undergoes a crossover: for strong interactions, the EPR can be accurately inferred by discarding the dynamics of internal states; for strong interactions, such an inference severely underestimates the EPR. In contrast with other works~\cite{Ramaswamy2018, Borthne2020, Giardina2022, yu2022energy, ferretti2024, boffi2024}, the EPR is highest (and equal to the non-interacting value) when the system is in the apolar phase: this result is consistent with another thermodynamically-consistent study where interactions systematically reduce the EPR~\cite{Agranov}. At the macroscopic level, the EPR admits a decomposition between diffusive and reactive contributions, as found in generic reaction-diffusion dynamics~\cite{Aslyamov2023, Avanzini2024}.

Our approach for formulating active lattice models with thermodynamic consistency can be generalized to a broader class of dynamics; for instance, to quantify the dissipation of clock models~\cite{chatterjee2020flocking, solon2022susceptibility}, dynamics with non-reciprocal interactions~\cite{martin2023exact, Loos2023, Rieger2023, dopierala2025, mohite2025}, or pulsating systems~\cite{Zhang2023, manacorda2024, banerjee2024}. Finally, similar thermodynamically consistent lattice models can potentially be used to properly evaluate the heat capacity of active systems~\cite{Maes2023}, in line with recent experiments in living systems~\cite{Foster2021, Foster2023}.


\acknowledgements{We acknowledge useful discussions with Tal Agranov, Michael E. Cates, and Robert L. Jack. This research was funded in part by the Luxembourg National Research Fund (FNR), grant references 14389168 and 14063202, and also grant no. NSF PHY-2309135 to the Kavli Institute for Theoretical Physics (KITP). ME is funded by the FNR CORE project ChemComplex (Grant No. C21/MS/16356329). 
}


\appendix

\section{Stochastic field dynamics}\label{app:action}

In this Appendix, we derive the stochastic dynamics of the fields for the local number of $(+,-)$ particles, respectively denoted by $p({\bf r},t)$ and $m({\bf r},t)$ [Eq.~\eqref{eq:field_def}]. To this end, we start with the master equation describing the microscopic lattice dynamics [Eq.~\eqref{eq:ME}], and coarse-grain for a small lattice spacing.

We consider the action functional $A$ that defines the path probability proportional to $e^{-A}$ for our lattice dynamics [Eq.~\eqref{eq:ME}]. Following standard methods~\cite{Thompson2011, agranov_exact_2021, agranov_entropy_2022, falasco2023macroscopic}, we obtain
\begin{widetext}
\begin{equation}\label{eq:action}
\begin{aligned}
  A &= \int dt \sum_{i,j} \bigg\{-\hat p_{i,j} \dot p_{i,j} -\hat m_{i,j} \dot m_{i,j} + {\cal W}(i,j,-|i,j,+) p_{i,j} \left(e^{-\hat p_{i,j}+\hat m_{i,j}} -1\right) + {\cal W}(i,j,+|i,j,-) m_{i,j} \left(e^{-\hat m_{i,j}+\hat p_{i,j}} -1\right)
  \\
  &\qquad\qquad\qquad + \sum_{k,l} \Big[ {\cal W}(k,l,+|i,j,+) p_{i,j} \left(e^{-\hat p_{i,j}+\hat p_{k,l}} -1\right) + {\cal W}(k,l,-|i,j,-) m_{i,j} \left(e^{-\hat m_{i,j}+\hat m_{k,l}} -1\right) \Big] \bigg\} ,
  \end{aligned}
\end{equation}
\end{widetext}
where $(\hat p_{i,j}, \hat m_{i,j})$ are auxiliary variables. The rate ${\cal W}(k,l,s'|i,j,s)$ corresponds to a transition of a particle with state $s\in(+,-)$ at the lattice site $(i,j)$ to a state $s'\in(+,-)$ at lattice site $(k,l)$. In practice, a transition changes either the internal state of the particle [Eq.~\eqref{eq:W2}] or the position of the particle from a given lattice site to its neighbor [Eq.~\eqref{eq:W1}].

In the continuum limit, we expand the terms in Eq.~\eqref{eq:action} for a small $a$. First, the auxiliary variables read
\begin{equation}\label{eq:expand}
	\hat p_{k,l}(t) \simeq \hat p({\bf r}_{ij},t) + a \hat \nabla \hat p({\bf r}_{ij},t) + \frac 1 2  (a \hat \nabla)^2 \hat p({\bf r}_{ij},t) ,
\end{equation}
and a similar expansion holds for $\hat m_{k,l}$. Second, the rates for transitions between sites [Eq.~\eqref{eq:W1}] can be written as
\begin{equation}
\begin{aligned}
    {\cal W}(k,l,+|i,j,+) &\simeq \gamma \left [1 - \frac{a}{2T} \left( \hat \nabla \frac{\delta E}{\delta p} + \hat{\bf f} \right) \right ] ,
    \\
    {\cal W}(k,l,-|i,j,-) &\simeq \gamma \left [1 - \frac{a}{2T} \left( \hat \nabla \frac{\delta E}{\delta m} - \hat {\bf f} \right) \right ] ,
 \end{aligned}
\end{equation}
with $(\hat \nabla, \hat {\bf f})$ the projections of $(\nabla,{\bf f})$, respectively, onto the unit vector pointing from $(i,j)$ to $(k,l)$, and the force ${\bf f}$ is defined in Eq.~\eqref{eq:f_scale}. Third, the rates for transitions between states [Eq.~\eqref{eq:W2}] simplify as
\begin{equation}\label{eq:expand_bis}
	{\cal W}(i,j,+|i,j,-) \simeq {\cal W}_- ,
	\quad
	{\cal W}(i,j,-|i,j,+) \simeq {\cal W}_+ ,
\end{equation}
where $({\cal W}_+,{\cal W}_-)$ are defined in Eq.~\eqref{eq:rate_scale}. Substituting these expansions [Eqs.~(\ref{eq:expand}-\ref{eq:expand_bis})] into the action $A$ [Eq.~\eqref{eq:action}], we deduce 
\begin{equation}
\begin{aligned}
  A &\propto \int dt \int_V d{\bf r} \bigg\{ - \hat p \partial_t p -\hat m\partial_t m
  \\
  &\quad + D \left[  \frac p T \left( -\nabla \frac{\delta E}{\delta p} - {\bf f} \right) \cdot \nabla \hat p + \hat p  \nabla^2 p  + p(\nabla \hat p)^2 \right]
  \\
  &\quad + D \left[ \frac m T \left(-\nabla  \frac{\delta E}{\delta m} + {\bf f} \right) \cdot \nabla \hat m + \hat m  \nabla^2 m  + m(\nabla \hat m)^2 \right]
  \\
	&\quad + {\cal W}_+ p \left(e^{-\hat p+\hat m} -1\right) + {\cal W}_- m \left(e^{-\hat m+\hat p} -1\right) \bigg\} ,
  \end{aligned}
\end{equation}
where the diffusion coefficient $D$ is defined in Eq.~\eqref{eq:f_scale}. The stochastic differential equations associated with the action $A$ can be written as
\begin{equation}\label{eq:SPDE}
\begin{aligned}
  \partial_t p &= D \nabla \cdot \left[ \frac{p}{T} \left( \nabla  \frac{\delta E}{\delta p} + {\bf f} \right) + \nabla p + {\boldsymbol\xi}_p \right] - {\cal N}_+ + {\cal N}_- ,
  \\
  \partial_t m &= D \nabla \cdot \left[ \frac{m}{T} \left( \nabla  \frac{\delta E}{\delta m} - {\bf f} \right) + \nabla m + {\boldsymbol\xi}_m \right] + {\cal N}_+ - {\cal N}_- ,
  \end{aligned}
\end{equation}
where $(\mathcal{N}_+,\mathcal{N}_-)$ are Poisson noises with respective rates proportional to $({\cal W}_{+},{\cal W}_-)$, and $({\boldsymbol\xi}_p,{\boldsymbol\xi}_m)$ are independent Gaussian white noises with zero mean and correlations given by
\begin{equation}
\begin{aligned}
	\langle \xi_{p\alpha}({\bf r},t) \xi_{p\beta}({\bf r}',t') \rangle &= \frac{2p}{DV} \delta_{\alpha\beta} \delta(t-t') \delta({\bf r}-{\bf r}') ,
	\\
	\langle \xi_{m\alpha}({\bf r},t) \xi_{m\beta}({\bf r}',t') \rangle &= \frac{2m}{DV} \delta_{\alpha\beta} \delta(t-t') \delta({\bf r}-{\bf r}') .
\end{aligned}
\end{equation}
For large system sizes ($V\to\infty$), the contribution from the fluctuating terms in Eq.~\eqref{eq:SPDE} is negligible, yielding the macroscopic field dynamics in Eq.~\eqref{eq:hydro}.


\bibliography{refs}

\end{document}